\documentclass[oneocolumn,preprintnumbers,amsmath,amssymb,nofootinbib]{revtex4}
\pdfoutput=1

\usepackage{amsmath,latexsym,amssymb,amsfonts}
\usepackage{graphicx,color}
\usepackage{bm}
\usepackage{gensymb}

\begin{document}	
\preprint{\parbox[b]{1in}{ \hbox{\tt PNUTP-22/A05}  }}
\title{\textbf{Trouble with geodesics in black-to-white hole bouncing scenarios}}
\author{
Deog Ki Hong$^{a,b}$\footnote{{\tt dkhong@pusan.ac.kr}},
Wei-Chen Lin$^{a,b}$\footnote{{\tt archennlin@gmail.com}}
and
Dong-han Yeom$^{a,c,d}$\footnote{{\tt innocent.yeom@gmail.com}}
}
\affiliation{
$^{a}$\footnotesize{Center for Cosmological Constant Problem, Pusan National University, Busan 46241, Republic of Korea}\\
$^{b}$\footnotesize{Department of Physics, Pusan National University, Busan 46241, Republic of Korea}\\
$^{c}$\footnotesize{Department of Physics Education, Pusan National University, Busan 46241, Republic of Korea}\\
$^{d}$\footnotesize{Research Center for Dielectric and Advanced Matter Physics, Pusan National University, Busan 46241, Republic of Korea}
}

\begin{abstract}
By utilizing the thin shell approximation, we investigate the behavior of radial timelike geodesics in a black hole to white hole bouncing scenario with a mass (de-)amplification relation. We show that those geodesics lose energy after crossing the transition surface if the white hole mass is less than the black hole mass and vice versa. That is, the bounded timelike radial geodesics become closer to the event horizon in the mass decreasing direction. We then show that by tracing a finite amount of bouncing cycles along the mass decreasing direction, all bounded radial geodesics can be squeezed into the range of the stretched horizon while the black hole and white hole are still massive. Those highly squeezed geodesics are problematic since there exists a Planck-scale blueshift between them and the regular infalling trajectories. We also discuss the possible implication and rescues.

\end{abstract}

\maketitle

\newpage

\tableofcontents

\section{Introduction}
  
Understanding and resolving the singularity inside the black hole is a fundamental and important task in modern theoretical physics. According to general relativity, as well as the singularity theorem \cite{Hawking:1970zqf}, it is inevitable to form a singularity as a result of gravitational collapse. At this singularity, our tools of the differential geometry all breaks down. Therefore, we need a new technique based on quantum gravity.

As there is no consensus about the final theory of quantum gravity, various approaches are proposed to resolve the singularity. Especially, in order to directly resolve the singularity, we need a non-perturbative approach toward quantum gravity \cite{DeWitt:1967yk}. We list two current \textit{attitudes} toward quantum gravity:
\begin{itemize}
\item[--] 1. \textit{Wave function}: Inside a black hole, we should introduce the wave function, and hence, inside a black hole is essentially quantum and there might be no classical analogy \cite{Bouhmadi-Lopez:2019kkt}.
\item[--] 2. \textit{Regular black holes}: Quantum gravitational corrections modify the Hamiltonian or the Lagrangian. As an effective classical solution of the modified theory, one can obtain a singularity-free solution that can be interpreted as a classical extension of spacetime \cite{Ayon-Beato:1999kuh}.  
\end{itemize}

One of the tantalizing approaches of the second attitude is to follow loop quantum gravity \cite{Gambini:2013ooa}. Based on this approach, one can find quantum corrections from loop representations of quantum states. Thanks to these corrections, we can modify the classical Hamiltonian; as a result, loop quantum modified black hole solutions can be different and even regular compared to the original classical black holes. Of course this is not the unique approach for regular black holes; there might be several regular black holes from vacuum bubbles \cite{Brahma:2019oal}, non-linear electrodynamics, phantom matters, or various modified gravity models \cite{Ayon-Beato:1999kuh}.

The question is, how can we remove the singularity. One of the traditional approaches is to substitute the singularity to an inner apparent horizon and a time-like regular boundary \cite{Bojowald:2018xxu}. However, this suffers from the instability issue of the Cauchy horizon. On the other hand, the loop quantum gravity inspired models prefer to rely on the time reversal symmetry \cite{Brahma:2021xjy}. In other words, it indicates that a black hole phase should be smoothly connected to a white hole phase once we include loop quantum gravity corrections. However, there might be three approaches for this construction:
\begin{itemize}
\item[--] According to Ashtekar and Bojowald \cite{Ashtekar:2005cj}, the black hole phase is connected to the white hole phase, where the future infinity is connected in one universe with quantum gravitational corrections only inside a black hole.
\item[--] According to Haggard and Rovelli \cite{Haggard:2014rza}, the black hole phase is connected to the white hole phase, where the future infinity is connected in one universe with quantum gravitational corrections that can reach outside a black hole \cite{Brahma:2018cgr}.
\item[--] According to Ashtekar, Olmedo and Singh \cite{Ashtekar:2018lag}, as well as Bodendorfer, Mele and Munch \cite{Bodendorfer2}, the black hole phase is connected to the white hole phase only inside the event horizon; hence, the future infinity of both phases are not connected in one universe.
\end{itemize}
Now the first and the second approaches are considered as a theoretically incomplete description \cite{Bojowald_criticism,Brahma:2018cgr}. There are some critical discussions of the third approach \cite{Bouhmadi-Lopez:2019hpp}, but it might be still self-consistent \cite{Bouhmadi-Lopez:2020oia}. However, it is fair to say that we need to check the theoretical consistency of the black-to-white hole bouncing models, for example, the stability of the white hole phase as the infalling particles penetrate the white hole region.

In this paper, we study the consistency of the black hole to the white hole bouncing models, especially by using geodesics. First, in Sec.~\ref{sec:mod}, we briefly review the loop quantum gravity based models, e.g., the Bodendorfer, Mele and Munch model \cite{Bodendorfer2}. This model can be approximately described by a thin-shell model that connects a black hole to a white hole with different mass parameters. This thin-shell model might be considered as a generalization of the black-to-white hole bouncing models. In Sec.~\ref{sec:geo}, We study geodesics in this background. We conclude that geodesics must be biased near the event horizon either future direction or past direction. If geodesics must be squeezed near the event horizon, the instability should increase along either the future direction or the past direction. We demonstrate this instability by showing that there exists a Planck-scale blueshift between those highly squeezed geodesics and the regular infalling trajectories. Thus, either the future or past instabilities strongly indicate the self-inconsistency of the completeness of the spacetime. Finally, in Sec.~\ref{sec:pos}, we discuss alternative ideas as well as possible future research topics.

\section{\label{sec:mod}Model}

\subsection{Bodendorfer-Mele-Munch model}

We first review a loop quantum gravity inspired black hole model which was proposed by Bodendorfer, Mele, and Munch \cite{Bodendorfer2}. We introduce the metric ansatz
\begin{equation}
d s_{\pm}^2= - \frac{a(r)}{L_{0}^{2}} dt^{2} +  \frac{n(r)}{a(r)} dr^{2} + b(r)^{2} d\Omega^{2},
\end{equation}
where $L_{0}$ is an infrared cut-off in the non-compact direction. Introducing canonical variables $v_{1} \equiv (2/3)b^{3}$,  $v_{2} \equiv 2 a b^{2}$, and corresponding canonical momenta $P_{1}$, $P_{2}$, we obtain the Hamiltonian density
\begin{equation}
\mathcal{H} = 3 v_{1} P_{1} P_{2} + v_{2} P_{2}^{2} - 2.
\end{equation}
By introducing the prescription of loop quantum gravity, i.e.,
\begin{equation}
P_{1,2} \rightarrow \frac{\sin \lambda_{1,2}P_{1,2}}{\lambda_{1,2}},
\end{equation}
one can obtain a regular black hole solution.

The effective solution explains a big-bounce near the putative spacelike singularity. We define the mass of the black hole phase $M_{-}$ and that of the white hole phase $M_{+}$. There is no fundamental relation between $M_{-}$ and $M_{+}$.\footnote{The mass amplification or deamplification relation was already observed in Ref. \cite{Corichi:2015xia}, where there is no fundamental relationship between mass parameters.} According to Bodendorfer-Mele-Munch, one needs to require the relation
\begin{equation}
M_{+} = M_{-} \left( \frac{M_{-}}{m} \right)^{\beta-1}
\end{equation}
to avoid indefinite mass amplification or de-amplification, where $m$ is a constant and $\beta = 5/3$ or $3/5$.

\subsection{Thin-shell generalization}

The Bodendorfer-Mele-Munch model is a consistent prescription \cite{Bouhmadi-Lopez:2020oia}, but technically the solution structure is complicated. In order to describe the mass amplification relation in a generic and consistent way, we introduce the thin-shell formalism \cite{Israel:1966rt}. We introduce the metric ansatz
\begin{equation}\label{Schwarzschild_metric_inside_BWH}
d s_{\pm}^2=-(-f_{\pm})^{-1}d r_{\pm}^2+(-f_{\pm})d t_{\pm}^2+r_{\pm}^2d\Omega^2, 
\end{equation}
which describes the metric inside the horizon,
\begin{equation}\label{f(r)_function}
f_{\pm} = 1-\frac{2M_{\pm}}{r} ,
\end{equation}
and $-$ denotes the black hole phase and $+$ denotes the white hole phase. The junction surface is at $r_{\pm} = r(\tau)$, where the induced metric along the space-like surface is given as
\begin{equation}
ds_{\mathrm{shell}}^{2} = d\tau^{2} + r^{2}(\tau) d\Omega^{2}.
\end{equation}
At the shell, the Einstein equation must be satisfied. First, we need to impose the Einstein equation or the so-called Israel junction equation. The junction equation requires the energy-momentum tensor of the shell; hence, second, we need the energy conservation relation for consistency. Therefore, the equations of motion of the shell are
\begin{eqnarray}
\epsilon_{+}\sqrt{\dot{r}^{2} - f_{+}} - \epsilon_{-}\sqrt{\dot{r}^{2} - f_{-}} &=& 4\pi r \sigma,\\
\dot{\sigma} &=& - 2\frac{\dot{r}}{r} \left( \sigma - \lambda \right),
\end{eqnarray}
where the dot denotes the derivative with respect to $\tau$; the first equation is the junction equation and the second equation is the energy conservation relation \cite{Brahma:2018cgr}. $\epsilon_{\pm} = + 1$ if $r$ increases along the outward normal direction; otherwise, $\epsilon_{\pm} = -1$. Therefore, for the black hole phase, $\epsilon_{-} = 1$, while for the white hole phase, $\epsilon_{+} = -1$. $\sigma$ is the tension and $\lambda$ is the pressure of the shell. One can assume the equation of state $\lambda_{i} = - w_{i}\sigma_{i}$ ($i=1,2,...$) to obtain
\begin{eqnarray}
\sigma(r) = \sum_{i} \frac{\sigma_{0i}}{r^{2(1+w_{i})}},
\end{eqnarray}
where $\sigma_{0i}$ are constants. By plugging this, the energy conservation relation is automatically satisfied and the junction equation is simplified to
\begin{eqnarray}
\dot{r}^{2} + V(r) = 0,
\end{eqnarray}
where the effective potential 
\begin{eqnarray}
V(r) = -f_{-}(r) - \frac{(f_{-}(r) - f_{+}(r) - 16 \pi^{2}\sigma^{2}(r) r^{2})^{2}}{64 \pi^{2} \sigma^{2}(r) r^{2}}\,.
\end{eqnarray}
\begin{figure}[h!]
	\centering
	\includegraphics[scale=0.6]{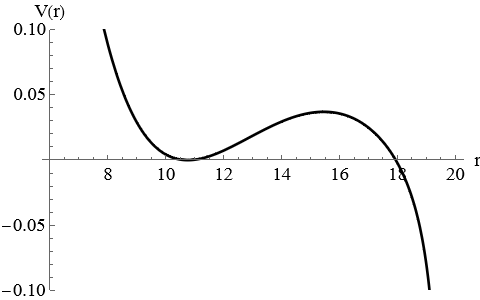}
	\caption{
	The effective potential $V(r)$ with $\sigma_{01} = 0.01496$, $\sigma_{02} = -0.3$, $w_{1} = -1$, $w_{2} = -0.5$, $M_{-} = 10$, and $M_{+} = 0.9 M_{-}$.
	}  
	\label{fig:exmple}
\end{figure}

\begin{figure}
\begin{center}
\includegraphics[scale=0.5]{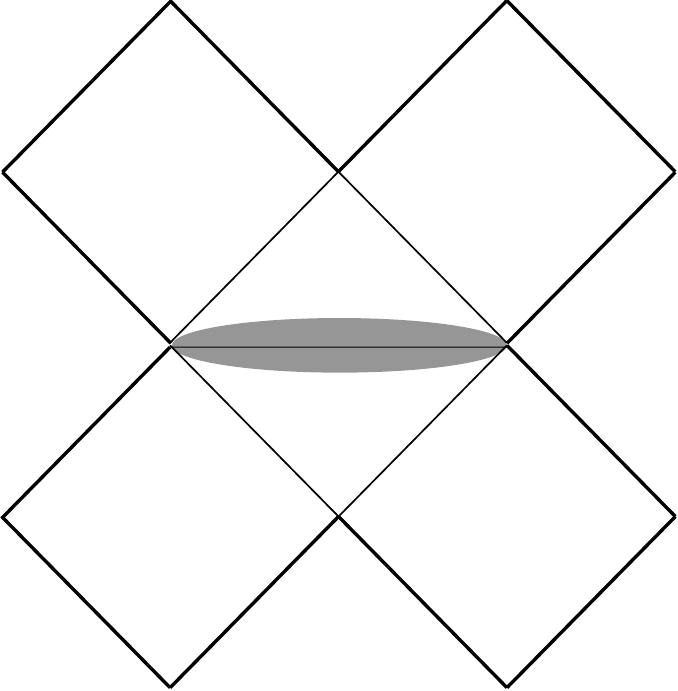}
\caption{\label{fig:penrose}The Penrose diagram of the black hole to white hole bouncing models, where the shell or quantum bouncing surface (gray colored region) is located around the classical singularity.}
\end{center}
\end{figure}

By tuning the parameters, it is possible to obtain a consistent model such that there exists $r_{0}$ at which $V(r_{0}) = V'(r_{0}) = 0$ and $V''(r_{0}) > 0$, needed for a static and stable shell ($r_0\approx 10.7$ in the model shown in Fig.~\ref{fig:exmple}). We need to tune the energy momentum tensor of the shell, but we assume that this can be justified from quantum gravitational effects. This thin-shell model effectively realizes several versions of the black to white hole bouncing models\footnote{It is worthwhile to mention that by utilizing the generalized Oppenheimer-Snyder model, the bouncing collapse have been investigated in general terms  \cite{BenAchour:2020bdt}. However notice that the resulting causal structure for a black-to-white hole bounce is different from what we consider here.  }. The final causal structure is shown in Fig.~\ref{fig:penrose}.

\section{\label{sec:geo}Geodesic analysis}

\subsection{\label{ssec:Energy}Energy Shift}

From this section on, we consider only the simplest scenario of the thin shell generalization, in which a thin shell is located at some fixed value of $r_{\pm}$. In such a case, from the first junction condition that the induced metric $d s^2_{\mathrm{shell}}$ on both sides of the thin shell must be the same, we  have 
\begin{equation}\label{induced_metric_relation_BWH}
(-f_{-})d t_{-}^2+r_{-}^2d\Omega^2=(-f_{+})d t_{+}^2+r_{+}^2d\Omega^2. 
\end{equation}
Thus, in an effective model that a black hole (BH) phase transition to a white hole (WH) phase at some minimal radius $b$, we must have
\begin{equation}\label{1JC_BWH_r=b}
r_{+}=r_{-}=b.
\end{equation}
and 
\begin{equation}\label{1JC_BWH_t_component}
\sqrt{-f_{-}(b)}d t_{-}=\sqrt{-f_{+}(b)}d t_{+},  
\end{equation}
which further leads to 
\begin{equation}\label{t_relation_BWH}
t_{+}=\sqrt{\frac{f_{-}(b)}{f_{+}(b)}}t_{-}. 
\end{equation}
Notice that it is a relation on the thin shell. 

Next, inside the event horizon, the four-velocity of a test particle moving along with a timelike radial geodesic in the BH phase is given by 
\begin{equation}\label{4_V_BH_phase}
\mathcal{U}_{-}^{\alpha}= \left( \Dot{r}_{-}, \Dot{t}_{-} \right)=\left( -\sqrt{E_{-}^2-f_{-}} \;, \frac{E_{-}}{f_{-}}\right), 
\end{equation}
while in the WH phase it has the form  
\begin{equation}\label{4_V_WH_phase}
\mathcal{U}_{+}^{\alpha}=\left( \Dot{r}_{+}, \Dot{t}_{+}\right)=\left( \sqrt{E_{+}^2-f_{+}} \;, \frac{E_{+}}{f_{+}}\right), 
\end{equation}
where $E_{\pm}$ is the energy density per unit mass of the test particle and the irrelevant angular components are suppressed. When $E_{\pm} \leq 1$, $E_{\pm}$ is related to the maximal radius $R_{\pm}$ the geodesic can reach by the relation $E_{\pm}^2=1-2M_{\pm}/R_{\pm}$.  We substitute the relation Eq.~(\ref{1JC_BWH_t_component}) into the t-component of Eq.~(\ref{4_V_WH_phase}) to have, at the thin shell,  
\begin{equation}\label{Blue_Shift_BWH_derive}
 \Dot{t}_{+}(b)=\frac{E_{+}}{f_{+}(b)}=\sqrt{\frac{f_{-}(b)}{f_{+}(b)}}\Dot{t}_{-}=\sqrt{\frac{f_{-}(b)}{f_{+}(b)}}\frac{E_{-}}{f_{-}(b)}, 
\end{equation}  
in which the $t$-component of Eq.~(\ref{4_V_BH_phase}) is used to obtain the last equality. Thus we have a relation of the energy shift as,  after the test particle crosses the thin shell,
\begin{equation}\label{Blue_shift_BWH}
E_{+}=\sqrt{\frac{f_{+}(b)}{f_{-}(b)}}E_{-}. 
\end{equation}

\begin{figure}[h!]
	\centering
	\includegraphics[scale=0.6]{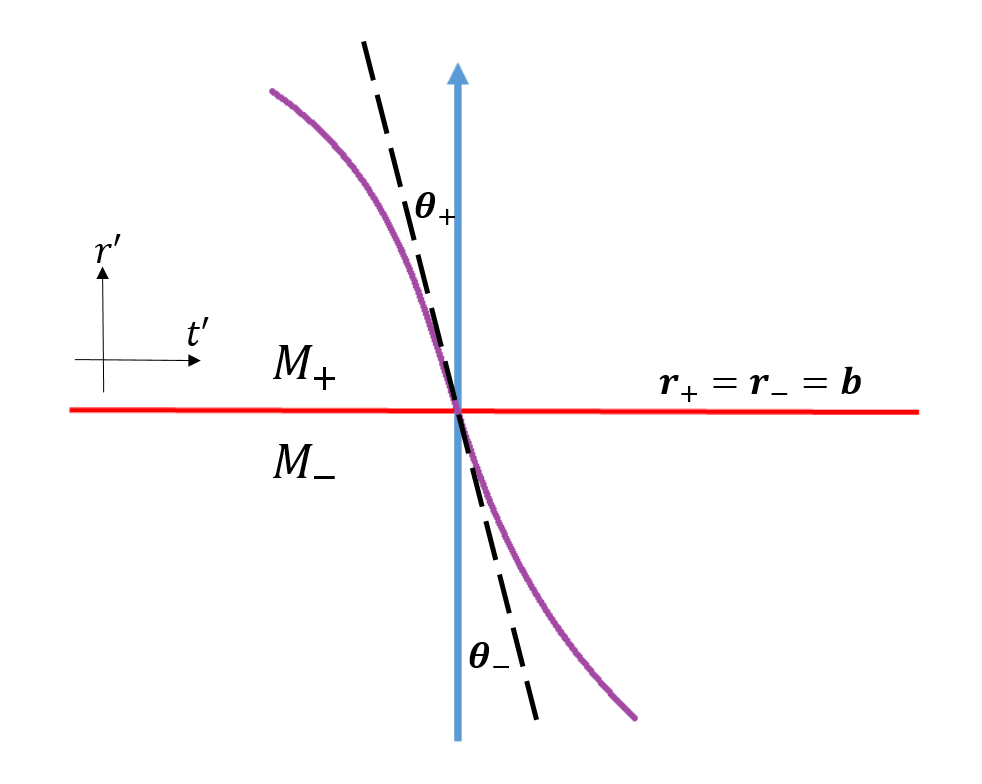}
	\caption{
	The intuitive equivalence between the smooth crossing of a geodesic (purple) and  $\theta_{+}= \theta_{-}$ shown in this picture is only true in a well-defined coordinate patch $(r', t')$ covering a open area which contains the entire thin shell.    
	}  
	\label{fig:No_Cusp}
\end{figure}

\begin{figure}[h!]
	\centering
	\includegraphics[scale=0.6]{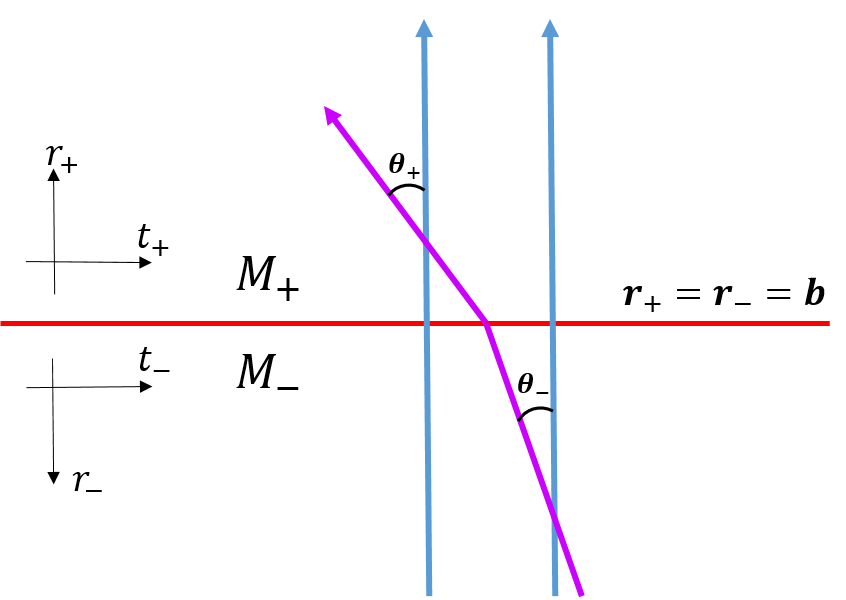}
	\caption{ An artificial cusp is generated for an arbitrary radial infalling geodesic (magenta) due to the fact that we just simply sew up two different coordinate patches at the thin shell. However, there is a special group of radial geodesics (light blue) which must be normal to the thin shell due to the fact that they always stay inside the B/W holes.}  
	\label{fig:No_Cusp_V2}
\end{figure}

Here we further show that this energy shift effect, Eq.~(\ref{Blue_shift_BWH}), is consistent with the condition that there is no cusp for any radial geodesic at the thin shell, that is, the trajectory of any geodesic must cross the thin shell smoothly.    

Before starting our argument, we would like to point out a caveat that one might naively think that the no cusp condition can be given by the following relation 
\begin{equation}\label{misleading_condition}
    \lim_{r_{+} \to b} \frac{\Dot{t}_{+}}{\Dot{r}_{+}}=\lim_{r_{-} \to b} \frac{\Dot{t}_{-}}{\Dot{r}_{-}}, 
\end{equation}
where $\Dot{r}_{\pm}$ and $\Dot{t}_{\pm}$ are given in Eqs.~(\ref{4_V_BH_phase}) and (\ref{4_V_WH_phase}). However, it is not true since we are using two coordinate patches and sewing them up right at the thin shell. So, Eq.~(\ref{misleading_condition}) does not make much sense, since each side of it explicitly depends on the vector components in different coordinates. To be able to use the continuity of the quantity $dt/dr$ as the condition for the no cusp for any geodesic at the thin shell, one needs to find a well-behaved coordinate patch covering an open region of the spacetime containing the whole thin shell, see Fig.~\ref{fig:No_Cusp}. However, we do not seek for such a coordinate patch in this work, but use a coordinate independent method by relating the no cusp condition to the continuity of a coordinate independent quantity at the thin shell. We then show that the continuity of this quantity leads to the energy shift relation Eq.~(\ref{Blue_shift_BWH}). Furthermore, we will see that by using this method, we can interpret the no cusp condition as a more physical condition that 
any timelike free falling observer sees no sudden change of the relative speed with respect to other timelike free falling observers after crossing the shell \footnote{To be more precisely, the relative speed is defined between a single free falling observer with some given $E_{-}=E_{a}$ and a family of free falling observers specified by a different energy density $E_{-}=E_{b}$. In this way, each point on the trajectory of the single observer intercepts with only one member of the family, so the relative speed between them can defined locally. See Ref. \cite{Borde:2001nh} for more discussion.}.

To begin with, we first notice that Eqs.~(\ref{4_V_BH_phase}) and (\ref{4_V_WH_phase}) contain a special case $E_{\pm}=0$, which leads to the four-velocity
\begin{equation}\label{4_velocity_KS_Observer}
\mathcal{V}_{\pm}^{\alpha}= \left( \Dot{r}_{\pm}, \Dot{t}_{\pm} \right)=\left(\pm \sqrt{-f_{\pm}}, 0 \right). 
\end{equation}
Thus the corresponding geodesics have a fixed $t_{\pm}-$component except the jump at the thin shell given by Eq.~(\ref{t_relation_BWH}). Here, we argue that this jumping of $t$-component should not cause discontinuity of the geodesics, and these group of geodesics are normal to the constant $r_{\pm}$ surfaces on both sides of the thin shell, see Fig.~\ref{fig:No_Cusp_V2}. Notice that this family of radial geodesics always stays inside the event horizon and the particles moving along with them can be thought as comoving with respect to the interior spacetime. Therefore, $\mathcal{V}_{\pm}^{\alpha}$ is the unique future-pointed unit four-vector normal to the thin shell since the four-velocity is always normalized as $-g_{\pm \alpha \beta}\mathcal{V}_{\pm}^{\alpha}\mathcal{V}_{\pm}^{\beta}=1$. 

Since $\mathcal{V}_{\pm}^{\alpha}$ is uniquely determined as mentioned above and any radial four-velocity $\mathcal{U}_{\pm}^{\alpha}$ has only a single degree of freedom due to normalization, the condition of no cusp for any radial geodesic crossing the thin shell is equivalent to the continuity of the inner product of  $\mathcal{U}_{\pm}^{\alpha}$ and $\mathcal{V}_{\pm}^{\alpha}$ at the thin shell as 
\begin{equation}\label{Gamma_matching_PM}
\gamma_{-}(b)=\gamma_{+}(b),
\end{equation}
where 
\begin{equation}\label{Gamma_limit_Def}
\gamma_{\pm}(b) \equiv \lim_{r_{\pm} \to b} -g_{\pm \alpha \beta}\mathcal{U}_{\pm}^{\alpha}\mathcal{V}_{\pm}^{\beta}. 
\end{equation}
Using Eqs.~(\ref{4_V_BH_phase}) and (\ref{4_V_WH_phase}) for $\mathcal{U}_{\pm}^{\alpha}$ and the metric Eq.~(\ref{Schwarzschild_metric_inside_BWH}), one can easily show that the condition Eq.~(\ref{Gamma_matching_PM}) leads to Eq.~(\ref{Blue_shift_BWH}). Lastly, notice that since we are considering only the timelike geodesics, the inner product  $\gamma=-g_{ \alpha \beta}\mathcal{U}^{\alpha}\mathcal{V}^{\beta}$ is in fact related to the relative speed $v_{rel}$ between $\mathcal{U}^{\alpha}$ and $\mathcal{V}^{\beta}$ as
\begin{equation}\label{relative_speed_gamma}
\gamma=\frac{1}{\sqrt{1-v_{rel}^2}}.
\end{equation}
The relation Eq.~(\ref{Gamma_matching_PM}) can be therefore reinterpreted as the statement that all timelike free falling observers see no sudden change in the relative speed with respect to other timelike free falling observers when crossing the shell. 

\subsection{\label{ssec:Max_R}Changing of the maximal radius reached by the test particles}

Now we discuss the consequence of the energy shift relation Eq.~(\ref{Blue_shift_BWH}). Especially, we  consider the limit $b \sim l_{Pl} \ll 2GM_{\pm}$, which corresponds to the usual assumption that the quantum gravity effect is important only near the Planck scale. And from here, we put the gravitational constant $G$ back for the convenience of later discussion. In such a limit, Eq.~(\ref{Blue_shift_BWH}) reduces to
\begin{equation}\label{energy_shift_in_Mass_approx}
E_{+}=\sqrt{\frac{M_{+}}{M_{-}}}E_{-}, 
\end{equation}
which, by using $E_{\pm}^2=1-2GM_{\pm}/R_{\pm}$, can be further rewritten as  
\begin{equation}
1-\frac{2GM_{+}}{R_{+}}=\frac{M_{+}}{M_{-}}\left(1-\frac{2GM_{-}}{R_{-}}\right). 
\end{equation}
Notice that having a negative $R_{\pm}$ is still a physical situation corresponding to $E_{\pm}>1$ for the energy density of the test particle.  The only thing changes is that $R_{\pm}$ loses the meaning as the maximal radius reached by the geodesic. In an usual maximally extended Schwarzschild solution, those geodesics are unbounded since they can reach the spatial infinity either in the distance future or past. However, the distinction between the bounded and unbounded radial geodesics changes in the BH/WH cyclic model as follows.   

By using the ratio $\alpha=M_{+}/M_{-}$ to replace the WH mass $M_{+}$, we have 
\begin{equation}\label{R_pm_relation_approx}
\frac{1}{R_{+}}=\frac{1}{R_{-}}+\frac{1-\alpha}{\alpha}\frac{1}{2GM_{-}}. 
\end{equation}
To understand this relation, we first assume $R_{+} \to \infty$, which is the condition that the test particle emitted by the white hole just has the enough energy to reach infinity without being sucked into the black hole again, \textit{i.e.} saturating the unbound condition.  
With $R_{+} \to \infty$, Eq.~(\ref{R_pm_relation_approx}) gives
\begin{equation}
\frac{R_{-}}{2GM_{-}}=\frac{\alpha}{\alpha-1}. 
\end{equation}
For $\alpha >1$, \textit{i.e.} the WH mass is greater than the BH mass, $R_{-}$ is positive finite. It means that before entering the black hole, the particle is ``bounded" by the black hole as it can only reach the maximal radius $R_{-}$, but after crossing the BH to WH transition surface, it is able to reach infinite to become unbounded.

Next, if the WH mass is less than the BH mass $\alpha=M_{+}/M_{-}<1$, Eq.~(\ref{R_pm_relation_approx}) can be rewritten in the terms of the ratio $\beta_{\pm}=R_{\pm}/2GM_{\pm}$ for the bounded geodesics with positive finite $R_{+}$ as 
\begin{equation}\label{R_pm_relation_in_Beta}
\beta_{+}=\frac{\beta_{-}}{(\beta_{-}-1)(1-\alpha)+1},
\end{equation}
which shows that $\beta_{+}< \beta_{-}$ when $\alpha=M_{+}/M_{-}<1$ by noticing $\beta_{\pm} >1 $. This means that a bounded geodesic will become closer to the next event horizon after crossing the BH/WH transition surface along the mass decreasing direction. We now discuss geodesics for the BH/WH cyclic model with a decreasing mass.

\subsection{\label{ssec:Squeezing}Squeezing of the bounded geodesics}

\subsubsection{Squeezing geodesics into the stretched horizon}

Now we consider a decreasing cycle with $\alpha=M_{i+1}/M_{i}<1$ starting from some reference stage with initial values $M_0$ and $R_0$, which satisfy conditions
$M_0 \gg m_{Pl}=G^{-1/2}$ and $R_0-2GM_0 \gg l_{Pl}=G^{1/2}$ respectively \footnote{We set $\hbar=c=1$, but keep $G$ to make the distinction between mass and length explicitly.}. Then at stage $n$ we have
\begin{equation}\label{mass_ladder}
M_n=\alpha^n M_0,
\end{equation}
and 
\begin{equation}\label{energy_ladder}
E^2_{n}=\alpha^n E^2_0, 
\end{equation}
where Eq.~(\ref{energy_shift_in_Mass_approx}) is used.
We then ask the question that at which stage the maximal radius of the geodesic falls into the range of the stretched horizon. That is, at certain stage labeled as $Y$, we have 
\begin{equation}\label{stretched_horizon_eq}
E^2_Y=1-\frac{2GM_Y}{2GM_Y+l_{Pl}} \approx \frac{G^{1/2}}{2GM_Y}, 
\end{equation}
where the condition $2GM_Y \gg l_{Pl}$ and the relation $l_{Pl}=G^{1/2}$ are used in the approximation. Notice that the condition $2GM_Y \gg l_{Pl}$ guarantees that at stage $Y$, the black and white holes can still be treated classically, that is, we still have $b \ll 2GM_{Y} $ so Eq. (\ref{energy_shift_in_Mass_approx}) holds. 
Using Eqs.~(\ref{mass_ladder}), (\ref{energy_ladder}) and (\ref{stretched_horizon_eq}), we have 
\begin{equation}\label{stretch_horizon_condition}
Y=\frac{\log( 2 \frac{M_0}{m_{Pl}}E_0^2)}{2(-\log \alpha)}, 
\end{equation}
where $m_{Pl}=G^{-1/2}$ is used. 

However, this relation might not be always true for any given reference values $M_0$ and $R_0$ since the mass of the B/W hole also decreases stage by stage. Therefore, we need to ensure that the condition $M_Y \gg  m_{Pl}$ still holds when all bounded geodesics are squeezed in the range of stretched horizon. Using, Eqs.~(\ref{mass_ladder}) and (\ref{stretch_horizon_condition}), this condition leads to the following relation between the initial parameters
\begin{equation}\label{classical_condition}
\frac{M_0}{m_{Pl}} \gg 2E^2_0. 
\end{equation}
Notice that the condition of a bounded geodesic is simply given by $E^2_0 < \alpha^{-1}$ when the approximation Eq.~(\ref{Blue_shift_BWH}) is valid. Eq.~(\ref{classical_condition}) shows that generally, for any reasonable choice of initial $M_0$ and $E_0$, the maximal radii reached by the bounded geodesics merge into the range of stretched horizon way before the BH/WH masses able to reach the Planck scale.  For instance, if in a decreasing model with $\alpha=1/2$ we choose $M_0$ to be the solar mass $M_0=M_{\odot} \sim 10^{38}m_{Pl}$  and $E_0$ saturating the limit for the bounded geodesics: $E_0^2=\alpha^{-1}=2$. Then by using above relations, we can see that at the stage when all bounded geodesics merge into the stretched horizon \cite{Thorne:1986iy}, the BH/WH masses is about $M_Y \sim 0.5 \times 10^{19}m_{Pl}$, which is still many orders above the Planck scale.

\begin{figure}[h!]
	\centering
	\includegraphics[scale=0.35]{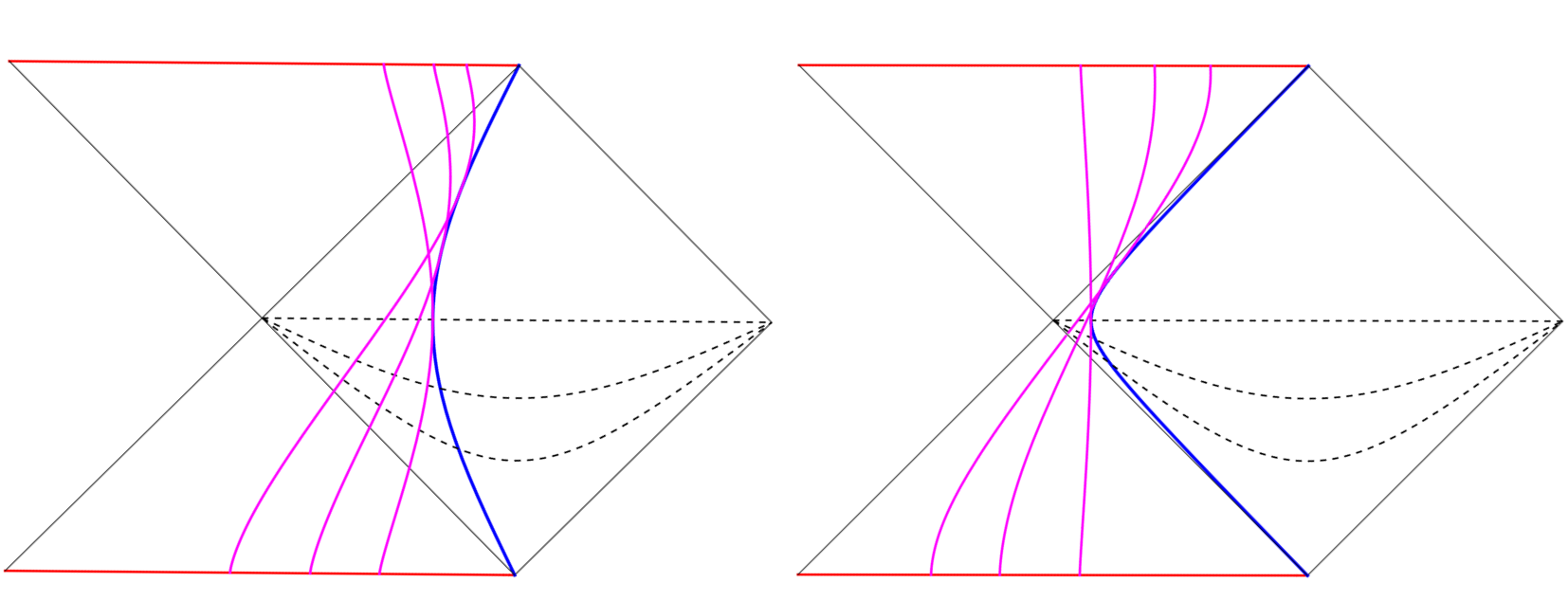}
	\caption{In this BH/WH cyclic model, a set of well-separated bounded geodesics in one stage (left) would be squeezed within a length shorter than the Planck distance either in some future or past stages (right) while the BH/WH masses are still much greater than the Planck mass. }  
	\label{fig:Squeezing geodesics}
\end{figure}

\subsubsection{Throwing massive particles into the BH/WH cyclic model}

Let us use a thought experiment to see what kind of bizarre scenario could occur if this type of BH/WH cyclic model exists. Firstly, for a decreasing mass model,  considering that at some stage, there is a distant observer, Alice, releasing one baseball per minute into the black hole, say for one hour. Then after those baseballs travel through a certain number of the BH/WH cycles, a second observer, Bob, at some future stage will see that all sixty baseballs are squeezed inside a Planck length range outside the black and white holes. If Alice measures the black hole mass to be around the solar mass $M_{\odot} \sim 10^{38}m_{Pl}$, then as estimated previously, the black hole mass for Bob will still be much larger than the mass of a baseball, which is about  $10^{7}m_{Pl}$. For instance, in the previous $M_{i+1}/M_{i}=0.5$ case, the black hole mass observed by Bob must be greater than $0.5 \times 10^{19}m_{Pl}$. Thus, those baseballs should not be able to disturb the spacetime structure significantly and are still valid to be treated as some classical test particles.\footnote{Indeed, the tidal force can easily tear those baseballs apart, but it  doesn't change the theoretical issue here, that is, a set of originally well-prepared system can evolve into a system involving Planckian scale.}  Also notice that since it only takes finite proper time for the baseballs to enter and then exit a BH/WH cycle, those baseballs would be seen in the squeezed phase after a finite amount of proper time experienced by them. So once we assume the existence of this type of decreasing mass BH/WH cyclic model, the squeezing scenario in principle must exist when the spacetime structure away from the bouncing surface still can be treated classically. Inversely, in a mass increasing model, if we trace back the history of a group of bounded particles ejected from a white hole, then at some earlier stage (universe), they could all be squeezed within the stretched horizon as viewed by an observer at that stage.

\subsection{\label{ssec:Planck_Accel}A Planck-scale accelerator}

Although well-separated radial geodesics can weirdly evolve into the squeezed geodesics in  BH/WH cyclic models with mass difference as shown previously, those squeezed geodesics alone might not cause issues. Since one can argue that when the test particles moving along with those bounded geodesics, interactions between them unavoidably happen. Thus, before the questions such as how those particles interact and where the energy comes from are answered, it is still undetermined whether the existence squeezed geodesics is problematic or not. Such scrutiny may be required if we only consider the squeezed geodesics alone. Here, we circumvent this complication involving a discussion of  different types of possible interactions. In the following, we demonstrate that the existence of the Planck scale squeezed geodesics indeed can cause instability when we consider the interaction between them and a group of regular infalling massive particles.

\begin{figure}[h!]
	\centering
	\includegraphics[scale=0.55]{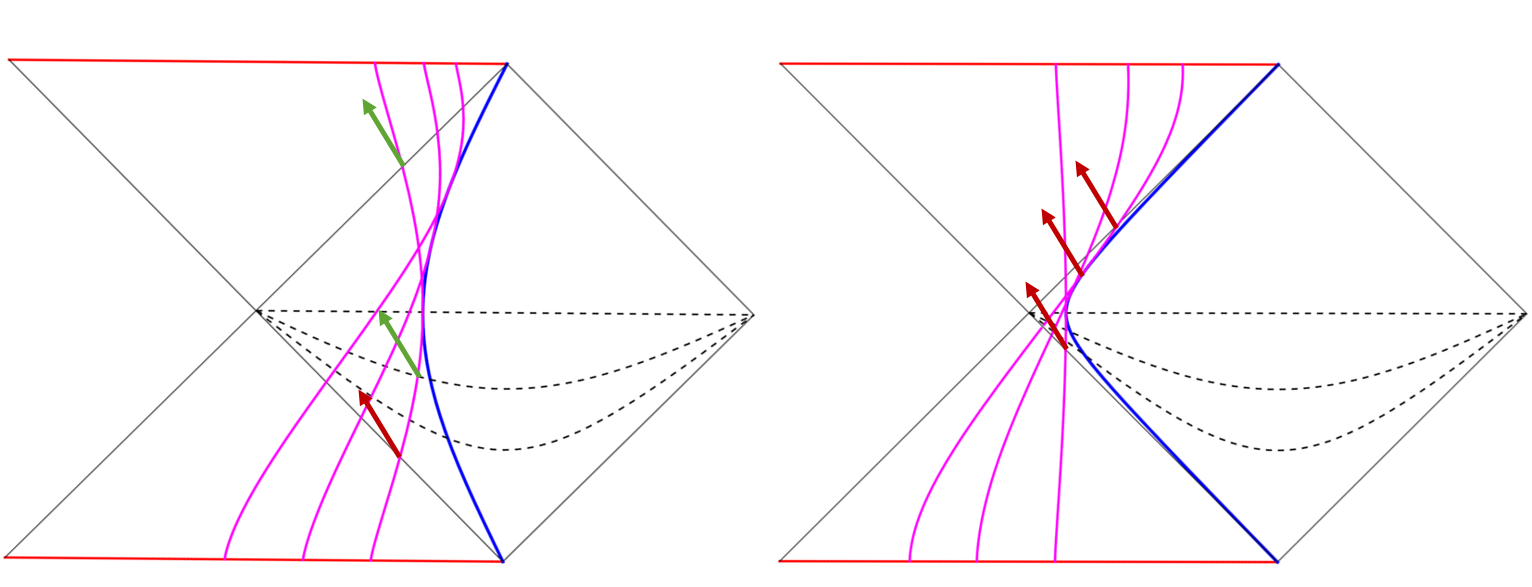}
	\caption{In this figure, arrows represent the four-velocities of radial infalling observers, whose trajectories do not have to follow geodesics. On the left Penrose diagram, different radial infalling observers encounter a group of regular bounded geodesics and measure the relative speed between them, and those arrows in green represent the situation that the relative speed falls into the regular range. The only observer (red arrow) measuring a relative speed extremely close to the speed of light is the one who encounters the group of bounded geodesics close to the WH event horizon. However, this issue is common if we agree with the existence of the white hole part of the maximally extended Schwarzschild solution. And we can argue that it is not really a physical issue since it is impossible to prepare such an observer moving along with the WH event horizon at a certain moment. On the right Penrose diagram, however, not only the above mentioned divergence at the WH event horizon exists, but there is a Planck scale blueshift of order $\sqrt{2M/m_{Pl}}$ to the relative speed between infalling observers and those highly squeezed geodesics. This energetic interaction between those two groups of particles indicates instability of the background spacetime.}  
	\label{fig:Planck_Accel}
\end{figure}

Let us consider the quantity $\gamma =-g_{\alpha \beta}\mathcal{U}^{\alpha}\mathcal{V}_{in}^{\beta}$ again, and focus on  $\gamma$ outside the event horizon. This time, we choose  $\mathcal{U}^{\alpha}$ to be the four-velocity of a bounded radial geodesic 
and  $\mathcal{V}_{in}^{\beta}$ to be the four-velocity of a radial infalling observer whose trajectory does not have to be a geodesic.
Nevertheless, at any instant, we can relate $\mathcal{V}_{in}^{\beta}$ to a radial infalling geodesic specified by some energy density $E'$.
Thus, for simplicity, we can use a radial infalling geodesic specified by some energy density $E' \sim \mathcal{O}(1)$ in the following discussion without losing generality. 
By using Eqs. (\ref{Schwarzschild_metric_inside_BWH}), (\ref{4_V_BH_phase}) and (\ref{4_V_WH_phase}), we have 
\begin{equation}\label{the_gamma_factor_general}
\gamma \equiv -g_{\alpha \beta}\mathcal{U}^{\alpha}\mathcal{V}_{in}^{\beta}=\frac{1}{f}\left[EE' \pm \sqrt{E^2-f}\sqrt{E'^2-f}\right], 
\end{equation}
where the plus sign on the r.h.s. is for the outgoing $\mathcal{U}^{\alpha}$ from the white hole, while the minus sign is for the infalling  $\mathcal{U}^{\alpha}$ to the black hole. 
We can further simplify the calculation without altering the qualitative result by setting  $E'=1$, which makes Eq. (\ref{the_gamma_factor_general}) reduce to 
\begin{equation}\label{gamma_E=1}
\gamma=\frac{1}{f}\left[E \pm \sqrt{\frac{2GM}{r}}\sqrt{E^2-f}\right]. 
\end{equation}
Notice that $\gamma$ is finite at the BH event horizon but divergent at the WH horizon. One can check this by substituting $r=2GM+\delta$ with $\delta \ll 2GM$ to get 
\begin{equation}\label{gamma_around_2M}
\gamma \approx \frac{1}{x}\left[E \pm E \left(1-\left(\frac{1}{2}+\frac{1}{2E^2}\right)x+ \left(\frac{3}{8}-\frac{1}{4E^2}+\frac{1}{8E^4}\right)x^2\right)\right],
\end{equation}
where $x \equiv \delta/(2GM)$ and $x \ll E^2$ is assumed in the above expansion. For the infalling  $\mathcal{U}^{\alpha}$ at the BH event horizon, Eq.~(\ref{gamma_around_2M}) leads to
\begin{equation}\label{gamma_at_2M}
\gamma (r=2GM) = \frac{E}{2}+\frac{1}{2E},  
\end{equation}
and notice that this is an exact relation for $E>0$. While for the outgoing $\mathcal{U}^{\alpha}$ around the WH event horizon, $\gamma$ diverges as
\begin{equation}\label{gamma_WH_around_2M}
\gamma (\delta \to 0) \approx \frac{4GME}{\delta}.
\end{equation}
This issue is common if we agree with the existence of the white hole as part of the maximally extended Schwarzschild solution. Thus, we can argue that Eq.~(\ref{gamma_WH_around_2M}) is not really a physical issue since it is impossible to prepare such an observer moving along with the WH event horizon at a certain moment. So the divergence of $\gamma$ at the white hole event horizon is not a real problem, see Fig. \ref{fig:Planck_Accel} (left). On the other hand, from Eq.~(\ref{gamma_at_2M}) for the infalling $\mathcal{U}^{\alpha}$, one can immediately see that $\gamma$ can be arbitrary large for those squeezed radial geodesics with $E \ll 1$.  This result is expected since those squeezed geodesics turn their direction from outgoing to infalling in the region close to the event horizon, the qualitative results from Eqs.~(\ref{gamma_at_2M}) and (\ref{gamma_WH_around_2M}) should be the same, see Fig. \ref{fig:Planck_Accel} (right).

To obtain a quantitative estimate, we choose $\mathcal{U}^{\alpha}$ to be the four-velocity of a squeezed geodesic whose maximal radius is at the boundary of stretched horizon, \textit{i.e.} $R=2GM+l_{Pl}$. By substituting $E=1-2GM/R$ and $R=2GM+l_{Pl}$ into Eq.~(\ref{gamma_E=1}) with $r=2GM+\delta$, we have the following relation 
\begin{equation}\label{gamma_small_E}
\gamma \approx \frac{2GM}{\delta}\left(\sqrt{\frac{l_{Pl}}{2GM}}\pm \sqrt{\frac{l_{Pl}-\delta}{2GM}} \right),
\end{equation}
where $-$ is again for the infalling $\mathcal{U}^{\alpha}$ and $+$ is for the outgoing $\mathcal{U}^{\alpha}$. Also notice that $l_{Pl}, \delta \ll 2GM$ is used and $\delta \leq l_{Pl}$. 
By using 
\begin{equation}\label{Planck_length_ratio}
0 \leq \frac{\delta}{l_{Pl}} \equiv y \leq 1, 
\end{equation}
Eq. (\ref{gamma_small_E}) for the infalling $\mathcal{U}^{\alpha}$ can be rewritten as
\begin{equation}\label{Planck_accelerated_gamma}
\gamma \approx \sqrt{\frac{2GM}{l_{Pl}}}\left(\frac{1}{y}-\sqrt{\frac{1-y}{y^2}}\right)=\sqrt{\frac{2M}{m_{Pl}}}\left(\frac{1}{y}-\sqrt{\frac{1-y}{y^2}}\right), 
\end{equation}
with
\begin{equation}\label{the_num_factor_for_accel_gamma}
\frac{1}{2} \leq \left(\frac{1}{y}-\sqrt{\frac{1-y}{y^2}}\right) \leq 1  \quad when \quad 0^{+} \leq  y \leq 1, 
\end{equation}
where $G^{1/2}=l_{Pl}=m^{-1}_{Pl}$ is used. Since  $\gamma$ is related to the relative speed by Eq.~(\ref{relative_speed_gamma}),  Eq.~(\ref{Planck_accelerated_gamma}) shows a Planck scale blueshift $\sqrt{ 2M/m_{Pl}}$ between the four-velocities of infalling squeezed geodesics $\mathcal{U}^{\alpha}$ and of infalling observers $\mathcal{V}_{in}^{\beta}$. One should also notice that the result in Eq.~(\ref{Planck_accelerated_gamma}) is derived by using the condition $R=2GM+l_{Pl}$; therefore, for the squeezed radial geodesics deeper inside the stretched horizon, the similar blueshift can be larger. This highly blueshifted relative speed between the squeezed radial geodesics and infalling objects (observers) can cause energetic collision. Especially, based on the discussion in the subsection \ref{ssec:Squeezing}, not only the factor $\sqrt{ 2M/m_{Pl}}$ can be large, but the mass of the particles can also be of several orders above the Planck mass. Thus, we conclude that this type of spacetime structure is unstable once we put the behavior of radial geodesics into consideration.             

This means that if the classical singularity of a Schwarzschild solution is replaced by some BH/WH bouncing with mass difference, which can also be effectively modeled by the thin shell approximation, the bounded geodesics then suffer from the  Planckian squeezing effect mentioned in subsection \ref{ssec:Squeezing}. Then, this effect further leads to a Planck-scale blueshift between the squeezed geodesics and regular infalling trajectories, which is quantified  by using the group of radial infalling geodesics specified by $E'=1$ in Eq.~(\ref{gamma_small_E}). This issue indicates an unexpected break down of the BH/WH cycles much earlier than the naive expected break down when the BH/WH masses reaching the Planck scale.

\section{\label{sec:pos}Possible rescues and Discussion}


Loop quantum gravity offers possible solution both to the singularity inside a black hole and to the beginning of the universe (Big Bang singularity). In the early universe aspect, potential cyclic cosmological models as alternatives to cosmological inflation are still under consideration. Particularly, a cyclic cosmological model with a non-zero average Hubble expansion rate is suggested to be able to solve the long standing entropy issue in the early universe theory \cite{Ijjas:2021zwv}. However, in such an early universe model, by looking backward in time, massive particles can be infinitely blueshifted within their own finite proper time, which indicates a breakdown of the cyclic phase \cite{Borde:2001nh,Kinney:2021imp}.

Similarly, in this BH/WH bouncing model with mass difference inspired by loop quantum gravity, massive particles can gain or lose energy when crossing the bouncing surface. One might be wondering if this type of model suffers the similar infinite blueshift issue as the special cyclic cosmological model mentioned above. 
Although indeed a massive particle can gain energy in a increasing mass BH/WH cyclic model, the particle cannot be infinitely blueshifted since once the energy is large enough, it simply escapes the black hole at certain stage and becomes unbounded. Therefore, no massive particle can be accelerated to reach speed of light by the BH/WH cycles so this particular model is safe from this UV aspect.       
On the contrary, the existence of event horizon causes a different type of issue when particles losing their energy. Assuming the usual junction condition, the radial infalling particles lose their energy quicker than the decreasing masses of the BH/WH cycles in the sense that  
all of the bounded geodesics would be squeezed into the stretched horizon within their finite proper time while the BH/WH are still massive. 
We then show that those highly squeezed geodesics are problematic by demonstrating a Planck-scale blueshift between them and the regular infalling trajectories. Indeed, one might argue that radial geodesics infinitely close to the event horizon also exist in the usual Schwarzschild solution. However, the very point here is that those geodesics can be evolved from well-separated radial geodesics in the BH/WH cyclic model with mass difference. This might indicate either an early breakdown of the BH/WH cyclic picture before the BH/WH masses become comparable to the Planck mass, or some inner inconsistency of the loop quantum gravity theory in which BH/WH cyclic models with mass difference exist.\footnote{Our approach is based on the consideration that the black hole and white hole phases of the classical Schwarzschild solutions with different masses can be glued together at some minimal radius as shown in Section II. B. So the results, especially the energy shifting after crossing, rely on the usage of the classical Schwarzschild solution and the first Israel junction condition, \textit{i.e.} the validity of modeling a BH/WH bounce by using the classical Schwarzschild solution and the thin shell approximation. On the other hand, however, breaking one of the above assumptions also means that the spacetime described by such a BH/WH bouncing model with mass (de-)amplification significantly deviates from that of a classical Schwarzschild solution in regions away from the bouncing point.}        

\begin{figure}
\begin{center}
\includegraphics[scale=0.07]{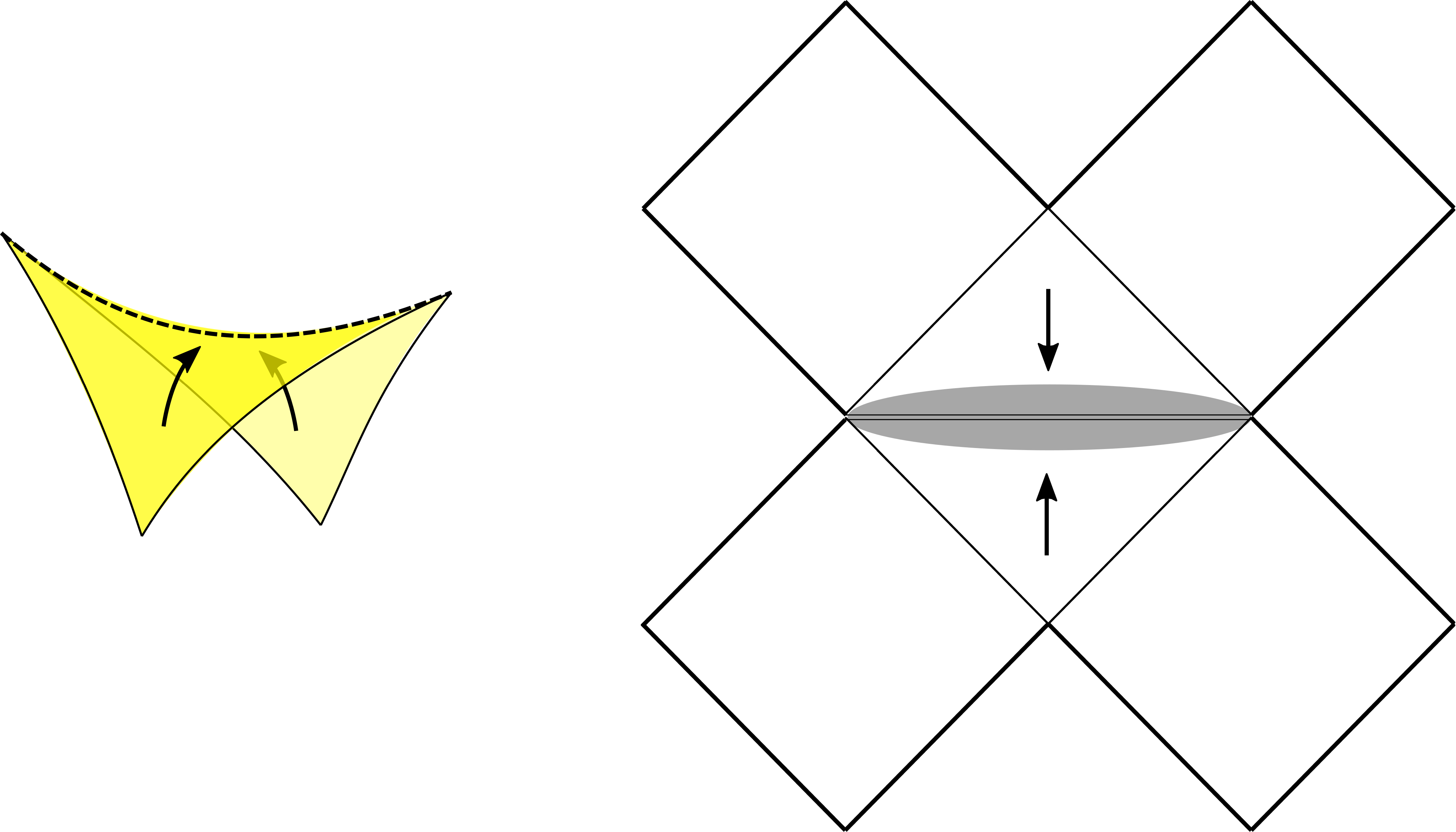}
\caption{\label{fig:AN}Annihilation-to-nothing interpretation \cite{Brahma:2021xjy}. Left: the black hole phase and the white hole phase are collide and annihilated. Right: as we define two arrows of time, one can interpret that two black hole phases are collided at the quantum bouncing surface.}
\end{center}
\end{figure}

If one wants to avoid such a potential instability, we may imagine two possibilities.
\begin{itemize}
\item[--] 1. If there exists a symmetry between $M_{+}$ and $M_{-}$, one may avoid the past or future squeezing. For example, if $M_{+} = M_{-}$ (perfectly symmetric \cite{Ashtekar:2018lag}) or $\beta = 3/5$ or $5/3$ (periodically symmetric \cite{Bodendorfer2}), the squeezing effect will be relaxed repeatedly. However, this interpretation requires the \textit{a priori} relation between $M_{+}$ and $M_{-}$; this is not very persuasive unless there exists a fundamental restriction to the mass parameters $M_{\pm}$.
\item[--] 2. The big bounce inside the black hole happens near the quantum gravitational regime. Around this quantum bouncing surface, there is no well defined arrow of time. So, there might be dual interpretations \cite{Bouhmadi-Lopez:2019kkt}; either we interpret there is one arrow only or there are two arrows. The latter interpretation is so-called the `annihilation-to-nothing' interpretation (Fig.~\ref{fig:AN}). In other words, there is no white hole phase, but there are two black hole phases, where these are annihilated near the singularity. If this is the case, we can definitely avoid all problems of the squeezing or mass amplification issues.
\end{itemize}
The latter interpretation provides a wisdom to understand generic quantum bouncing models of loop quantum gravity \cite{Brahma:2018elv}. Also, this provides the DeWitt boundary condition inside a black hole horizon that is recently emphasized in the literature \cite{Perry:2021mch}.

Lastly, can the squeezing effect be avoided if we consider the interaction between the test particles moving along with the squeezed geodesics? Is there any modification required to describe interaction in such a squeezing scenario? Or, for instance, we can consider a situation that two squeezed particles interact, in which one of them is able to gain enough energy to escape the black hole. Then, where does the other particle go? And also, can such an interaction cause some other unphysical issues? In this work, we merely point out the existence of the geodesic squeezing phenomenon in a generalized black-to-white hole bouncing scenario with mass difference, and then identify one problem in this type of model by considering the relation of those squeezed geodesics and other regular infalling trajectories. To answer above mentioned questions requires further studies and will be our future works. Especially, a detailed construction of the trajectories of radial geodesics might be an interesting future research direction.

\newpage

\section*{Acknowledgment}
This work was supported by the National Research Foundation of Korea (NRF) grant funded by the Korea government (MSIT) (2021R1A4A5031460). This work is also supported by Basic Science Research Program through the National Research Foundation of Korea (NRF) funded by the Ministry of Education (NRF-2017R1D1A1B06033701) (DKH) and 
by the National Research Foundation of Korea (Grant no.:2021R1C1C1008622) (DY).


\end{document}